\documentclass[aps,prb,reprint,a4paper,floatfix,groupedaddress,amsmath,amsfonts,amssymb]{revtex4-1}

\pdfoutput=1

\usepackage{mathptmx}
\usepackage[utf8]{inputenc} 
\usepackage{graphicx}
\usepackage{textcomp}
\usepackage{mathrsfs}
\usepackage{sidecap}
\usepackage[
,textwidth=17.5cm
,textheight=23.5cm
,verbose
,dvips
]{geometry}

\begin{document}

\title{Sub-meV linewidth in GaN nanowire ensembles: \\ 
absence of surface excitons due to the field-ionization of donors}

\author{Pierre Corfdir}
\email{corfdir@pdi-berlin.de}
\author{Johannes K. Zettler}
\author{Christian Hauswald}
\author{Sergio Fernández-Garrido}
\author{Oliver Brandt}
\affiliation{Paul-Drude-Institut für Festkörperelektronik, Hausvogteiplatz 5–7, 10117 Berlin, Germany}
\author{Pierre Lefebvre}
\affiliation{CNRS, Laboratoire Charles Coulomb, UMR 5221, 34095 Montpellier, France}
\affiliation{Laboratoire Charles Coulomb, Université Montpellier 2, UMR 5221, 34095 Montpellier, France}

\date{\today}

\begin{abstract}
We observe unusually narrow donor-bound exciton transitions (400~\textmu eV) in the photoluminescence spectra of GaN nanowire ensembles grown on Si(111) substrates at very high ($>850$°C) temperatures. The spectra of these samples reveal a prominent transition of excitons bound to neutral Si impurities which is not observed for samples grown under standard conditions. Motivated by these experimental results, we investigate theoretically the impact of surface-induced internal electric fields on the binding energy of donors by a combined Monte Carlo and envelope function approach. We obtain the ranges of doping and diameter for which the potential is well described using the Poisson equation, where one assumes a spatially homogeneous distribution of dopants. Our calculations also show that surface donors in nanowires with a diameter smaller than 100~nm are ionized when the surface electric field is larger than about 10~kV/cm, corresponding to a doping level higher than $2 \times 10^{16}$~cm$^{-3}$. This result explains the experimental observation: since the (D$^{+}$,X) complex is not stable in GaN, surface-donor-bound excitons do not contribute to the photoluminescence spectra of GaN nanowires above a certain doping level, and the linewidth reflects the actual structural perfection of the nanowire ensemble.
\end{abstract}


\maketitle

\section{Introduction}

In contrast to epitaxial layers, single-crystal GaN can be grown on Si substrates as well as on amorphous substrates in the form of nanowires with diameters ranging typically between 30 and 100~nm.\cite{Calleja2000,Consonni2011,Sobanska2014} The high crystal quality of GaN nanowires facilitates the investigation of fundamental aspects of these nanostructures by purely optical means, such as the role of the surface on their spontaneous emission.\cite{Schlager2008,Corfdir2009,Park2009,Brandt2010,Pfueller2010a,Gorgis2012,Korona2012} These studies have shown that the nanowire surface may affect both the radiative and nonradiative recombination processes of excitons. 

Concerning the latter process, surface recombination has been reported to be the dominant recombination process in GaN nanowires with a diameter of 30 nm or thinner \cite{Gorgis2012} despite the fact that the dangling bond states for the nonpolar surfaces of GaN are situated far from midgap.\cite{Segev2006,Lymperakis2013} Regarding the former process, both the wavefunction and energy of electrons and excitons bound to point defects are altered in the vicinity of a surface.\cite{Levine1965,Satpathy1983,Corfdir2012} This surface-induced change of the properties of point defects leads to a distribution of the emission energy of donor-bound excitons in nanowires\cite{Corfdir2009,Brandt2010} and to a modification of the lifetimes of the different radiative transitions involved as compared to the bulk case.\cite{Corfdir2009} The pinning of the Fermi level at the sidewalls of nanowires\cite{Segev2006,Lymperakis2013} adds further complexity to nanowire-based systems. As shown in Ref.~\onlinecite{Calarco2005}, this pinning induces radial electric fields within the nanowires and is responsible for their electrical depletion. 

The consequences of these surface potentials are manifold. For example, they modify the radiative decay rate of excitons\cite{Pfueller2010a,Lefebvre2012} and enhance the coupling between free and bound excitons with profound consequences for the exciton decay dynamics.\cite{Hauswald2013} In (In,Ga)N/GaN nanowire heterostructures, the interplay between the polarization and surface potentials may lead to a radial separation of electron and holes, resulting in a dramatic decrease of the internal quantum efficiency in the blue spectral range.\cite{Marquardt2013} 

To understand these phenomena, a detailed knowledge of the distribution and the magnitude of surface-induced electric fields as a function of the doping level (N$_\text{D}$) and the nanowire diameter ($\phi$) is indispensable. Furthermore, we need to understand the transition between the single impurity limit for nanowires with a low background doping and small diameters,\cite{Pfueller2010b,Gorgis2012} and the bulk-like case of a homogeneous dopant distribution reached for intentionally doped\cite{Calarco2005} or very large nanowires.\cite{Schlager2008}

In this paper, we present experimental results demonstrating that very high growth temperatures can induce the incorporation of Si into GaN nanowires grown on Si substrates, but can simultaneously result in a sub-meV linewidth of the donor-bound exciton transition for the ensemble. To understand these results, we investigate the role of the surface on the properties of donors in GaN nanowires theoretically. Using a combination of Monte-Carlo and envelope function calculations, we examine the validity of assuming a parabolic potential across the section of nanowires. We compute the binding energy of donors in the presence of surface-induced electric fields and discuss the doping- and diameter ranges for which neutral surface donors are stable and for which they are not. We find that neutral surface donors capable of binding excitons do not exist already for moderate doping levels (for example, $2 \times 10^{16}$~cm$^{-3}$ for a nanowire diameter of 100~nm), which explains the record linewidths observed experimentally.  

\begin{figure*}
\includegraphics[scale=1]{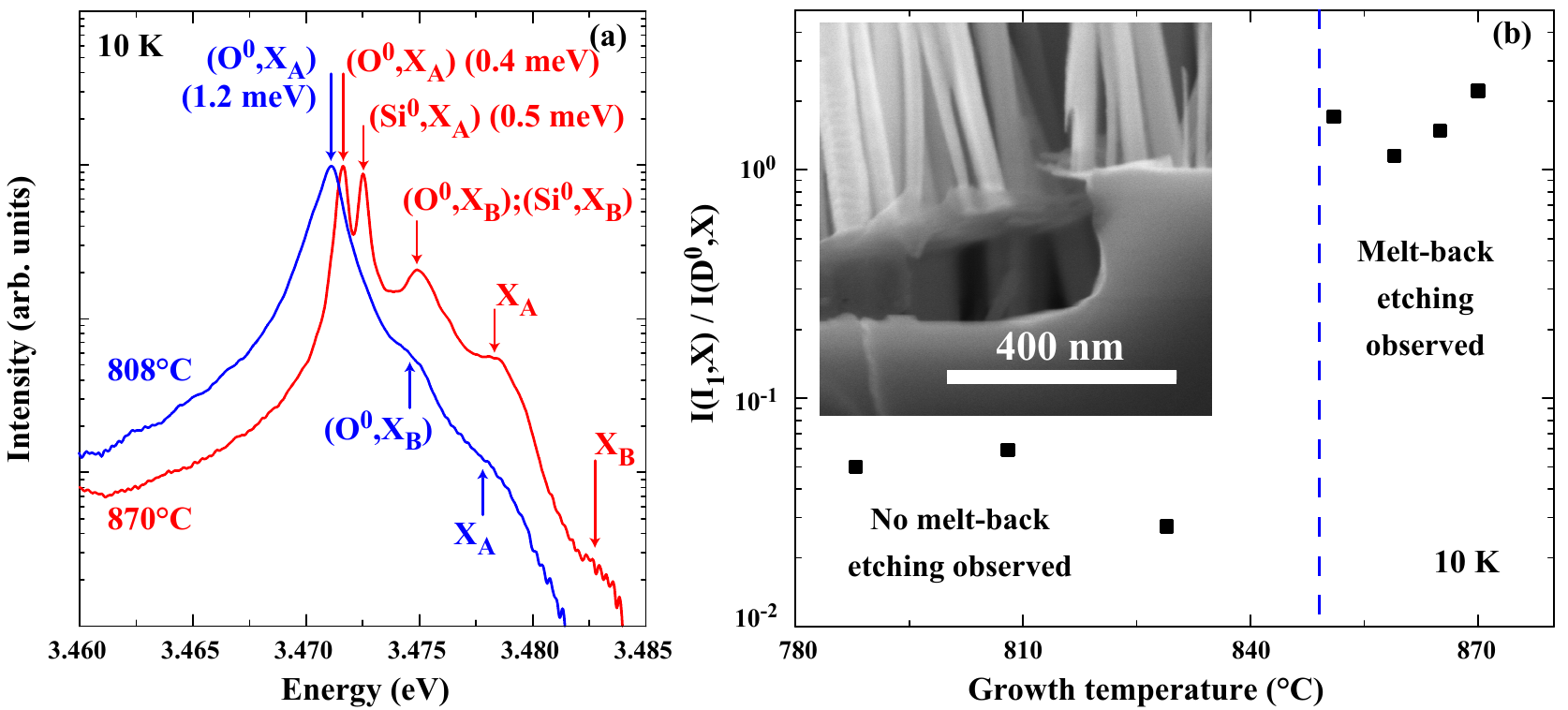}
\caption {(color online): (a) Low-temperature (10\,K) photoluminescence spectra of GaN nanowire ensembles grown at 808 and 870$^\circ$C on Si(111) and normalized to the (O$^0$,X$_\text{A}$) transition. The values in parentheses indicate the full-width at half maximum of the lines related to donor-bound A excitons. (b) Evolution of the intensity ratio of the transitions from excitons bound to I$_{1}$ stacking faults and neutral donors [I$_{(\text{I}_{1},\text{X})}$/I$_{(\text{D}^{0},\text{X})}$] at 10~K with growth temperature. The dashed line indicates the growth temperature above which Si melt-back etching is systematically observed. The inset shows a cross-sectional scanning electron micrograph of a pit caused by the melt-back etching of the Si substrate.}
\label{fig:FigurePLandSEM}
\end{figure*} 

The paper is organized as follows. Section \ref{sec:Methods} is devoted to the experimental methods used in this work. In Section \ref{sec:PL}, the photoluminescence spectra of GaN nanowires grown on Si at a temperature of 870$^\circ$C are presented. In Section \ref{sec:ParabolicPotential}, we present Monte-Carlo simulations which provide the range of values for $\phi$ and $N_D$ for which it is valid to approximate the radial potential across a nanowire by a parabola. In Section \ref{sec:BindingEnergy}, we calculate the binding energy of donors located at the nanowire surface, in the presence of surface-induced electric fields. In Section \ref{sec:Discussion}, we compare the results of our calculations with the experiments in Section \ref{sec:PL}. Finally, we present our conclusions in Section \ref{sec:Conclusion}.

\section{Experimental details}
\label{sec:Methods}

The nanowire ensembles studied here formed spontaneously during plasma-assisted molecular beam epitaxy of GaN on Si(111) substrates.\cite{Bertness2008,Fernandez-Garrido2009,Fernandez-Garrido2012,Consonni2013} The as-received substrates were etched in diluted (5\%) HF for 2~min. Prior to growth, the substrates were annealed in the growth chamber for 30~min at 880\,°C. After this process, the $7 \times 7$ surface reconstruction characteristic of a clean Si(111) surface appeared upon cooling to temperatures below 860\,°C. Subsequently, the substrate temperature was set to the desired value for growth, and the substrate was exposed to the N plasma for 10~min before opening the Ga shutter to induce the spontaneous formation of GaN nanowires. A series of samples was prepared using growth temperatures between 785 and 870\,°C. For all samples the active N flux provided by the radio frequency N plasma source was kept constant at $7.7 \times 10^{14}$~s$^{-1}$cm$^{-2}$. In contrast, the impinging Ga flux provided by a solid-source effusion cell was increased from $1.8 \times 10^{14}$ to $3.1 \times 10^{15}$~ s$^{-1}$\,cm$^{-2}$ to compensate for the exponential increase in Ga desorption with increasing temperature.\cite{Heying2000} Due to the high Ga desorption rate, the actual growth conditions were invariably N-rich as required for the spontaneous formation of GaN nanowires by molecular beam epitaxy.\cite{Fernandez-Garrido2013} The nanowires thus obtained have similar average lengths (2--3~\textmu m), mean diameters (70–-130~nm), densities ($5 \times 10^{9}$~cm$^{-2}$) and coalescence degrees.\cite{Brandt2014}

Photoluminescence spectroscopy was performed by exciting these GaN nanowire ensembles with the 325~nm line of a continuous-wave HeCd laser. The laser beam was attenuated with neutral density filters to a power of about 15~nW and was focused to a 1~\textmu m-spot using a near-UV objective with a numerical aperture of 0.45. We estimate that about 40 nanowires are excited simultaneously. The PL signal was collected using the same objective and dispersed by a monochromator with 80~cm focal length and a grating with 2400 lines/mm. For the current experiment, the spectral resolution was set to 0.25~\AA (i.e. about 250 ~\textmu eV). The dispersed signal was detected by a UV-sensitized liquid nitrogen-cooled charge coupled device.

\section{Donor-bound exciton transitions from a G\MakeLowercase{a}N nanowire ensemble with sub-\MakeLowercase{me}v linewidth}
\label{sec:PL}

\begin{SCfigure*}
\centering
\includegraphics[scale=1]{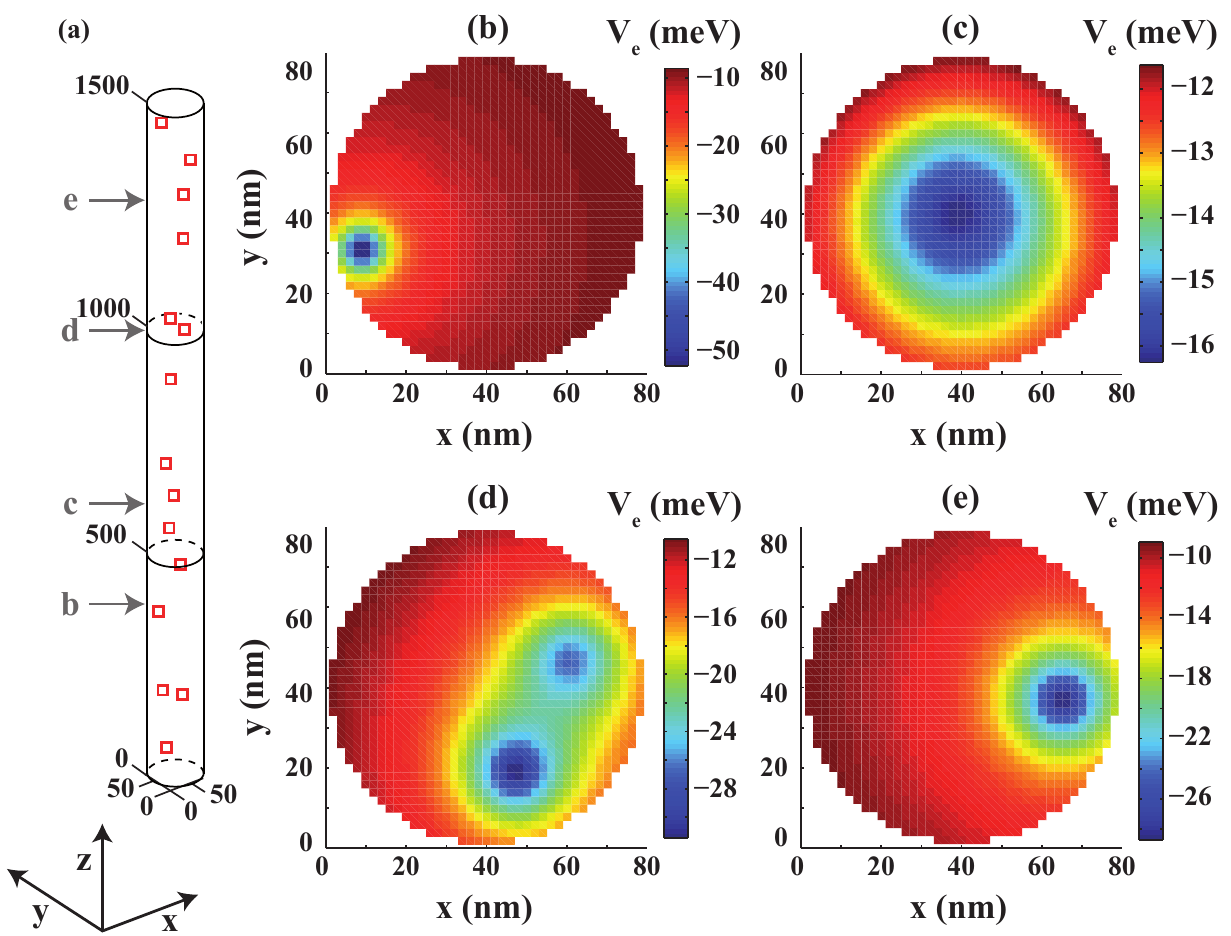}
\caption {(color online): (a) Cylindrical GaN nanowire with a diameter of 80~nm and a length of 1.5~\textmu m. The dimensions are to scale. Donors are distributed randomly (red squares) and exhibit an average concentration of about $2 \times 10^{15}$~cm$^{-3}$. The grey arrows indicate the positions along the nanowire length where the sections shown in panels (b)--(e) have been taken. (b)--(e) Potential across the nanowire section for various positions along the nanowire length. The potential $V_e$ is color-coded, with low and high potential values displayed in blue and red, respectively.}
\label{fig:FigurePotential}
\end{SCfigure*} 

Figure \ref{fig:FigurePLandSEM}(a) shows the low-temperature (10~K) photoluminescence spectrum of two ensembles of GaN nanowires formed at a growth temperature of 808 and 870\,°C. The spectrum of the nanowire sample grown at 808\,°C is dominated by a line centered at 3.4711~eV originating from the recombination of A excitons bound to neutral O donors on N sites [(O$^0$,X$_\text{A}$)]. The shoulders on the high-energy side of the (O$^0$,X$_\text{A}$) line are due to the emission from B excitons bound to neutral O donors [(O$^0$,X$_\text{B}$)] at 3.4743~eV and from free A excitons (X$_\text{A}$) at 3.4775~eV. The energy position of the free A exciton confirms that the net strain in our nanowire ensemble is virtually zero.\cite{Calleja2000,Robins2007,Corfdir2009,Park2009,Brandt2010} Despite this fact, the (O$^0$,X$_\text{A}$) line exhibits a full-width at half-maximum of 1.2~meV. This broadening arises from both the microstrain introduced by nanowire coalescence\cite{Jenichen2011,Kaganer2012} and the modification of the energy and the wavefunction of neutral donors and donor-bound excitons by the surface, as discussed in Refs.~\onlinecite{Corfdir2009,Brandt2010}.

Figure \ref{fig:FigurePLandSEM}(b) displays a cross-sectional scanning electron micrograph of the interface between the GaN nanowires and the Si substrate for a sample grown at a temperature of 870\,°C. A common phenomenon during the high-temperature growth of GaN on Si is the melt-back etching of the Si substrate.\cite{Zhu2013} This etching process arises from the formation of a Ga-Si eutectic alloy and results in the creation of large pits in the substrate. For our sample series, we observed these pits for growth temperatures higher than 850\,°C. 

The photoluminescence spectrum of the sample shown in Fig.~\ref{fig:FigurePLandSEM}(b) is compared to that of the sample grown below this critical temperature in Fig.~\ref{fig:FigurePLandSEM}(a). For all samples grown above 850\,°C, an additional line is observed at 3.4725~eV. This line corresponds in energy to the recombination of A excitons bound to neutral Si donors on Ga sites [(Si$^0$,X$_\text{A}$)], suggesting that the melt-back etching of the Si substrate is accompanied by the incorporation of Si in the nanowires. The incorporation of Si also manifests itself by an increase in the emission intensity of excitons bound to I$_{1}$ basal-plane stacking faults as shown in Fig.~\ref{fig:FigurePLandSEM}(b)]. In fact, the formation energy for stacking faults in GaN is reduced with increasing Si concentration as demonstrated theoretically by \citet{Chisholm2000,Chisholm2001}. 

Finally, we observe a significant reduction in the linewidth of the  donor-bound exciton lines for all samples grown above 850\,°C. For the nanowire ensemble grown at 870$^\circ$C, full-width at half-maxima of 400 and 500~$\mu$eV are measured for the (O$^0$,X$_\text{A}$) and (Si$^0$,X$_\text{A}$) lines, respectively. All samples grown at or above 850$^\circ$C exhibit this correlation between the occurence of Si meltback-etching, a strong (Si$^0$,X$_\text{A}$) transition, and sub-meV donor-bound exciton linewidths. This finding suggests that the higher donor concentration caused by the additional incorporation of Si is responsible for the narrow linewidth of our high-temperature observed for high growth temperature. To understand this apparently paradoxical result, we investigate theoretically in Sections \ref{sec:ParabolicPotential} and \ref{sec:BindingEnergy} the properties of donors in GaN nanowires as a function of the nanowire diameter and dopant concentration.

\section{Surface-induced parabolic potential}
\label{sec:ParabolicPotential}

\begin{figure*}[htbp]
\includegraphics[scale=1]{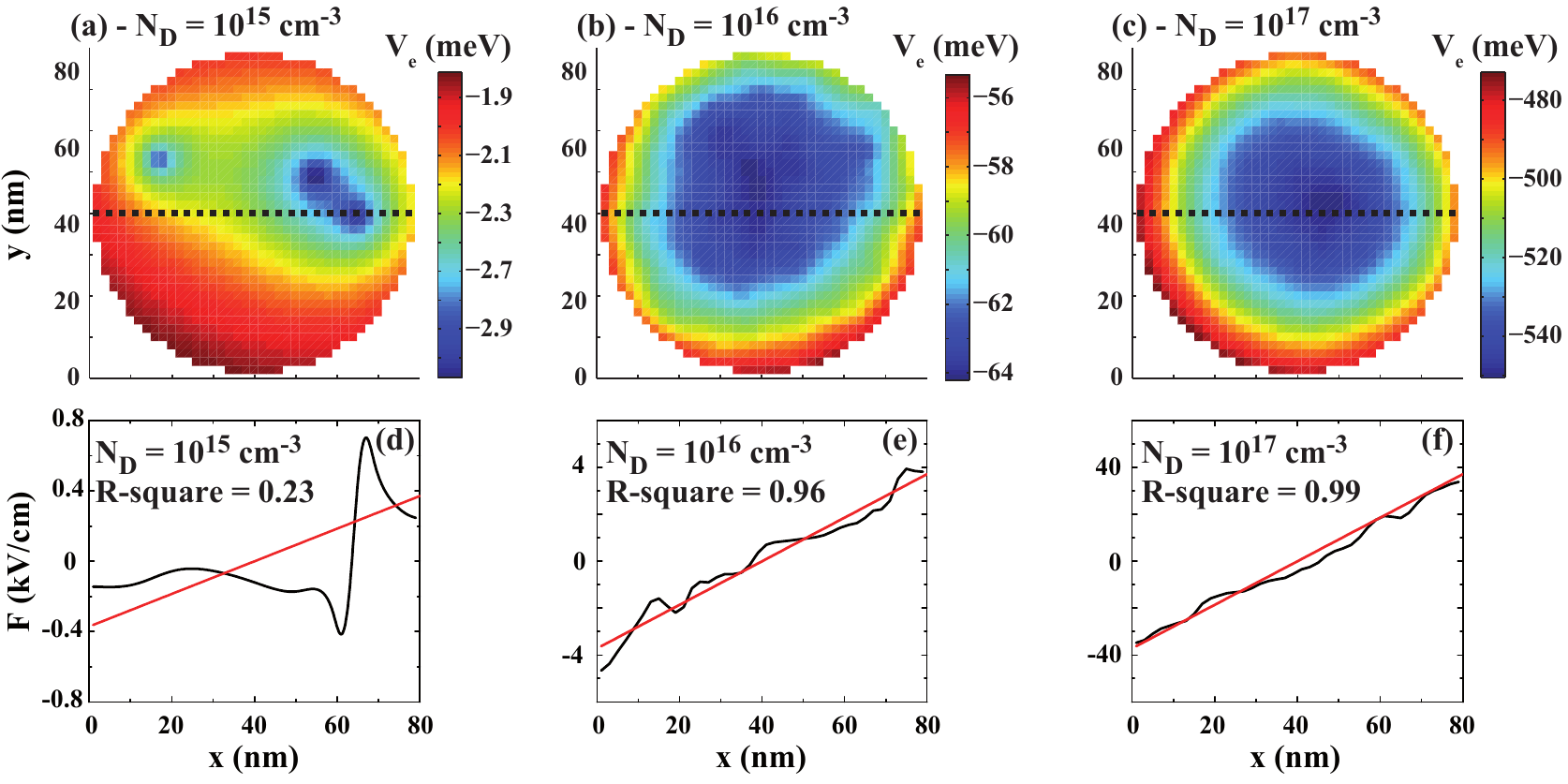}
\caption {(color online): (a)--(c): Average surface potential for a nanowire with a diameter and a length of 80 nm and 1.5 $\mu$m, respectively. The average doping concentration $N_D$ is (a) 10$^{15}$, (b) 10$^{16}$, and (c) 10$^{17}$~cm$^{-3}$. The nanowires simulated in (a), (b) and (c) exhibit 3, 84 and 709 donors, respectively. The potential $V_e$ is color-coded, with low and high potential values coded in blue and red, respectively. (d)--(f) Corresponding electric field along the $x$-axis (black line) across the center of the nanowire as indicated by the horizontal black dashed line in (a)--(c). The red line shows the result of a fit by Eq.~\ref{eqn:fit}. The R-square value characterizes the deviation between the fit and the computed electric field.}
\label{fig:FigurePotAverage}
\end{figure*} 

In this section, we employ Monte-Carlo simulations to evaluate the potential in nanowires with a diameter $\phi$ between 20 and 200~nm and a donor concentration $N_D$ between $10^{15}$ and $10^{18}$~cm$^{-3}$. The donors are distributed randomly in the nanowires and we assume that the nanowires are fully depleted. Figure~\ref{fig:FigurePotential}(a) shows one realization for the distribution of donors in a nanowire with $\phi = 80$~nm, $L = 1.5$~\textmu m and $N_D = 2\times10^{15}$~cm$^{-3}$. The simulated nanowire contains 15 donors. 

Figures~\ref{fig:FigurePotential}(b)--\ref{fig:FigurePotential}(e)
show the potential across various cross-sections of the nanowire. The shape of the potential as well as the magnitude of the electric field across the section of the nanowire fluctuates along the nanowire length. For such a low doping density, each individual impurity strongly affects the local potential felt by the electron. In transport experiments such as those reported in Refs.~\onlinecite{Calarco2005,Sanford2010}, the electronic properties of a single nanowire are averaged along the nanowire length. Accordingly,  Figs.~\ref{fig:FigurePotAverage}(a)--\ref{fig:FigurePotAverage}(c) display the surface potential averaged along the nanowire axis [the $z$-axis in Fig.~\ref{fig:FigurePotential}(a)] for a nanowire with $\phi = 80$~nm and $N_D$ equal to 10$^{15}$, 10$^{16}$ and 10$^{17}$~cm$^{-3}$. Figures~\ref{fig:FigurePotAverage}(d)--\ref{fig:FigurePotAverage}(e) show the corresponding electric field along the $x$-axis [dashed line in Figs.~\ref{fig:FigurePotAverage}(a)--\ref{fig:FigurePotAverage}(c)] compared to that expected from the Poisson equation for a homogeneous distribution of charges with a concentration equal to $N_D$ in an infinitely long cylindrical nanowire with diameter $\phi$. For $N_D$ = 10$^{17}$~cm$^{-3}$, we obtain a good agreement between the computed $V_e$ and the prediction of the Poisson equation. For donor concentrations equal to and larger than this value, the donor distribution can thus be considered as homogeneous. In contrast, for $N_D$ = 10$^{15}$~cm$^{-3}$, there is no longer any correlation between $V_e$ and the straight line predicted by the Poisson equation. For low donor concentrations well below $N_D$ = 10$^{16}$~cm$^{-3}$, the shape of $V_e$ is sensitive to the detailed distribution of donors and the surface-induced potential can in no way be reproduced by a parabola.

For obtaining a quantitative criterion for the validity of the assumption of a parabolic potential for each set of values $(N_D,\phi)$, we fit the evolution of the electric field across the section of the nanowire with the following linear regression:

\begin{equation}
\label{eqn:fit}
F(x)=F_\text{surf} \left(\frac{2x}{\phi}-1 \right),
\end{equation}

where $F_\text{surf}$ is the electric field at the surface of the nanowire and $0<x<\phi$ is the position along the $x$-axis. We impose the condition that the electric field on the axis of the nanowire ($x=\phi / 2$) is equal to zero for symmetry reasons. Note that Eq.~\ref{eqn:fit} applies only when the nanowire is fully depleted. For $\phi > 100$~nm, donors situated in the inner core of GaN nanowires remain most probably unionized\cite{Calarco2005} and $F$ can be derived using the expressions given in Ref.~\onlinecite{Sanford2010} or in Ref.~\onlinecite{Dobrokhotov2006}. We note, however, that there is no significant deviation between the results obtained by Eq.~\ref{eqn:fit} and the expressions in Refs. \onlinecite{Sanford2010,Dobrokhotov2006} when $\phi < 200$~nm.  

For a depleted nanowire and when surface potentials can be described by the Poisson equation, the electric field $F_\text{surf}^\text{Poisson}$ at the surface of the nanowire is:

\begin{equation}
\label{eqn:FsurfTh}
F_\text{surf}^\text{Poisson}=\frac{qN_D\phi}{4\epsilon},
\end{equation}

where $\epsilon = 9.5 \epsilon_0$ is GaN dielectric constant. To grade the quality of the fit obtained using Eq.~\ref{eqn:fit}, we utilize the R-square metric.\cite{Jin2001} The R-square is given by $1-\frac{SS_\text{res}}{SS_\text{tot}}$, where $SS_\text{res}$ and $SS_\text{tot}$ are the residual sum of squares and the total sum of squares, respectively. When the R-square value averaged over ten realizations of the nanowire is larger than 0.9, we consider that the surface potential in a nanowire with a given set ($N_D$,$\phi$) can be described by a parabola. 

\begin{figure}[t]
\includegraphics[scale=1]{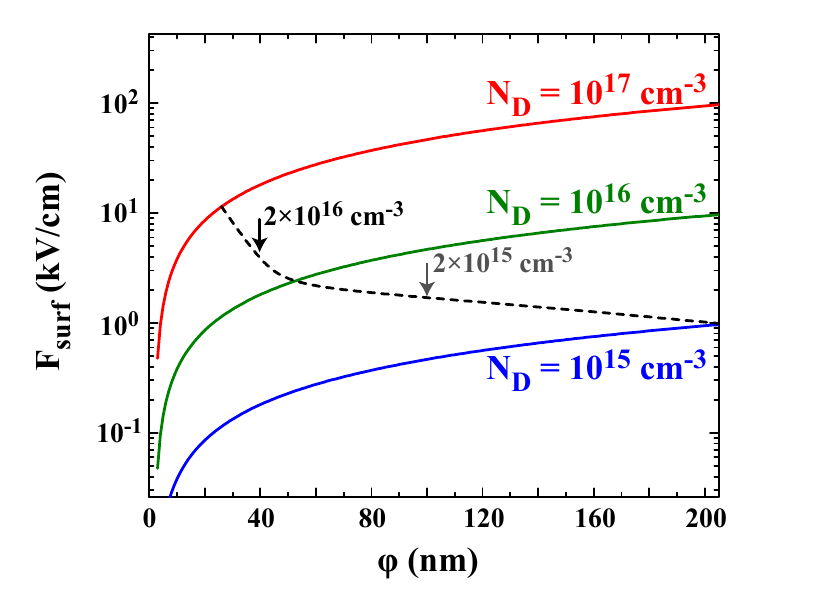}
\caption {(color online): Surface electric field $F_\text{surf}$ as a function of the nanowire diameter $\phi$ obtained using the Poisson equation for donor concentrations $N_D$ of $10^{15}$, $10^{16}$ and $10^{17}$~cm$^{-3}$ (blue, green and red solid lines, respectively). These values are reliable only above the black dashed line, which thus shows the lower limit for a given set of $\phi$ and $N_D$ for which the surface potential can be considered to be parabolic. The black and grey arrows indicate the $\phi$ below which the surface potential is not parabolic when $N_D= 2 \times 10^{16}$~cm$^{-3}$ and $2 \times 10^{15}$~cm$^{-3}$, respectively.}\label{fig:Summary}
\end{figure}

Our findings are summarized in Fig.~\ref{fig:Summary}, which shows 
$F_\text{surf}$ as a function of $\phi$ predicted by the Poisson equation for donor densities between $10^{15}$ and $10^{17}$~cm$^{-3}$. Below the dashed line, the R-square values are below 0.9 and the actual surface potential in the nanowire does not agree with the result of the Poisson equation for a homogeneous distribution of donors. When $\phi$ is decreased from 200 to 20~nm, $N_D$ has to be increased from about $10^{15}$ to a value in excess of $10^{17}$~cm$^{-3}$ for surface potentials in the nanowire to be correctly described by the Poisson equation. For a nanowire length of 1.5~$\mu$m, this condition corresponds to having more than 200 donors per nanowire. 

The experimental determination of $N_D$ is a challenging task, and  consequently only a few values have been reported.\cite{Sanford2010,Pfueller2010b} \citet{Pfueller2010b} obtained a value of $8\times 10^{15}$~cm$^{-3}$ from a statistical analysis of the exciton emission of single GaN nanowires with a diameter below 30~nm. This optical method is not affected by the considerations above. On the other hand, \citet{Sanford2010} deduced values between $5\times 10^{14}$ and $10^{15}$ cm$^{-3}$ for nanowires with $\phi$ ranging from 140 to 700~nm from an analysis of photoconductivity experiments based on the validity of the Poisson equation.\cite{Sanford2010} For these low donor concentrations, it is clear from Fig.~\ref{fig:Summary} that Eq.~\ref{eqn:fit} (or similar expressions such as those in Refs.~\onlinecite{Sanford2010,Dobrokhotov2006}) fails to reproduce the surface potential in nanowires with $\phi <$ 200 nm. It is, however, difficult to anticipate whether this breakdown of the continuum model leads to an over- or an underestimate of $N_{D}$ in such experiments.

\section{Binding energy of donors in the presence of internal electric fields}
\label{sec:BindingEnergy}

\begin{figure}
\includegraphics[scale=1]{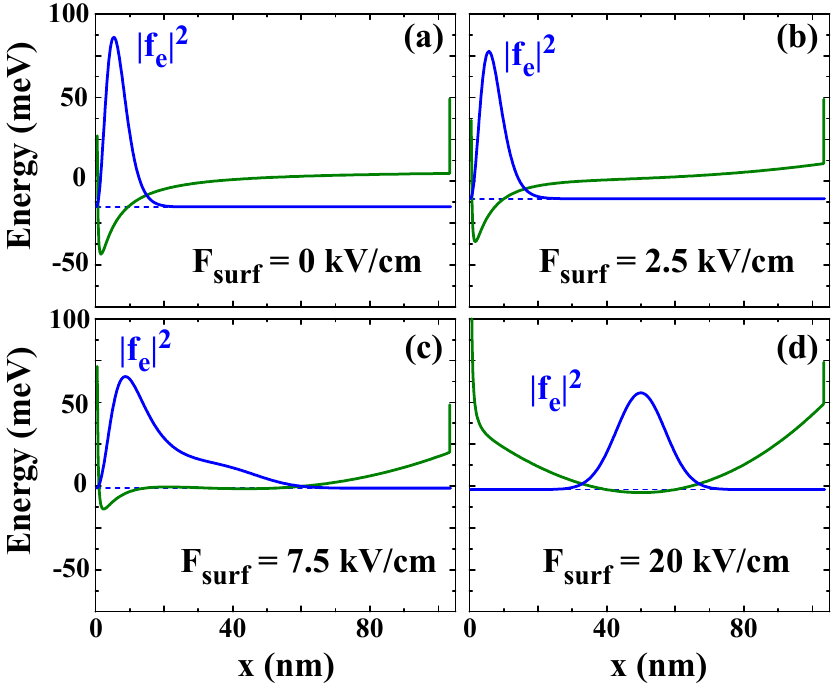}
\caption {(color online): Effective potential (green solid line) and probability distribution $|f_e|^2$ (blue solid line) for an electron in a GaN nanoslab with a thickness of 100~nm and for a surface electric field $F_\text{surf}$ of 0 (a), 2.5 (b), 7.5 (c) and 20~kV/cm (d). The donor is located at the left surface of the nanoslab. The y-scale for $|f_e|^2$ is shown in arbitrary units.}
\label{fig:FigureWavefunction}
\end{figure}

Photoluminescence spectroscopy is the most sensitive method for the detection of impurities in semiconductors. For the case of GaN nanowires, the high sensitivity of photoluminescence spectroscopy  has been exemplified in Ref.~\onlinecite{Pfueller2010b}, which reports the detection of the (D$^{0}$,X) transition even in single nanowires containing, on average, only a single donor. Photoluminescence spectroscopy is, however, completely insensitive to the presence of ionized donors. Because of the specific ratio of the effective masses of electrons and holes, the exciton-ionized donor-complex is not stable in GaN,\cite{Suffczynski1967} i.\,e., the presence of ionized donors does not lead to a radiative transition  observable in the photoluminescence spectra of GaN. Indirect evidence for the coexistence of neutral and ionized donors even at low temperatures has been obtained in the experiments reported in Refs.~\onlinecite{Pfueller2010a,Lefebvre2012}.

It is intuitively clear that the electric field associated with the ionized donors affects the energy of electrons bound to neutral donors. For flat-band conditions, the presence of the surface\cite{Levine1965,Satpathy1983} together with the dielectric mismatch\cite{Diarra2007,Pierre2010} leads to a strong site-dependence of the binding energy of electrons to donors in nanowires.\cite{Corfdir2012} In particular, the binding energy of donors at the surface of a thick GaN nanostructures is approximately 0.6 times that of donors in bulk.\cite{Corfdir2012} Since the magnitude of surface-induced electric fields is much stronger at the surface than in the core of nanowires (cf.\ Fig.~\ref{fig:FigurePotAverage}), the properties of neutral donors at the surface of GaN nanowires should be affected. 

Hence, we next calculate the binding energy of electrons to donors in nanowires in the presence of a surface-induced parabolic potential. We reduce the three-dimensional problem of a donor in a nanowire to a one-dimensional one by considering the simplified problem of a donor in a slab of GaN bounded by air. As shown in Ref.~\onlinecite{Corfdir2012}, the properties of donors in a nanowire with a diameter $\phi$ are well reproduced using a nanoslab geometry with a nanoslab thickness $d=\phi$. This fact remains true even when $\phi$ is as small as 10~nm. 

Regarding the symmetry of the potential for an electron in a nanoslab and in the presence of a donor, we choose to set up the problem using cylindrical coordinates. The Hamiltonian $\mathscr{H}$ for an electron with coordinates $(\rho,\theta,x)$ in a nanoslab and in the presence of a donor is given by:

\begin{equation}
\label{eqn:hamiltonian}
\mathscr{H} = -\frac{\hbar^2}{2m_e}\nabla^2 + V_e(x) + V_e^\text{d}(x)+V_e^\text{self}(x),
\end{equation}

where $m_e=0.2 m_0$ is the electron effective mass. In Eq.~\ref{eqn:hamiltonian},

\begin{equation}
\label{eqn:hamiltonian2}
V_e^\text{d} = -\frac{e^2}{4\pi\epsilon}
 \sum_{n=-\infty}^{n=+\infty} \left(\frac{\epsilon_r-1}{\epsilon_r+1}\right)^{|n|} \frac{1}{\sqrt{(x-d_n)^2+\rho^2}}   ,
\end{equation}

describes the interaction of the electron with the donor and the image charges of the donor while

\begin{equation}
\label{eqn:hamiltonian3}
V_e^\text{self} =\frac{e^2}{4\pi\epsilon}
 \sum_{n=-\infty,n \neq 0}^{n=+\infty} \frac{1}{2}\left(\frac{\epsilon_r-1}{\epsilon_r+1}\right)^{|n|}\frac{1}{|x-x_n|} .
\end{equation}

accounts for the electron self-energy. The donor and its images are on the $x$-axis with coordinates $x=d_0$ and $x=d_{n,n \neq 0}$, respectively, and $x_n$ corresponds to the on-axis coordinate of the $n^{th}$ image of the electron. The coordinates $d_n$ and $x_n$ are obtained following Ref.~\onlinecite{Kumagai1989}. The potential term $V_e(x_e)$ is the potential across the nanowire section and we assume it to be parabolic. We choose the following trial wavefunction for the electron: 

\begin{equation}
\label{eqn:trialwavefunction}
\Psi_e(x,\rho) = \frac{N}{\lambda}f_e(x)e^{-\frac{\rho}{\lambda}}
\end{equation}

with $N$ being a normalization factor, $\lambda$ describing the extent of the electron wavefunction in the plane of the nanoslab and $f_e(x)$ representing the electron envelope function along the confinement axis. The Schrödinger equation is solved numerically using the effective potential formalism (for more details, see Refs. \onlinecite{Corfdir2009b,Corfdir2012}) allowing us to deduce the electron wavefunction and binding energy for various values of $F_\text{surf}$ and $\phi$. The electron density giving rise to such a surface electric field in nanowires can be obtained using Eq.~\ref{eqn:FsurfTh}.

\begin{figure}[b]
\includegraphics[scale=1]{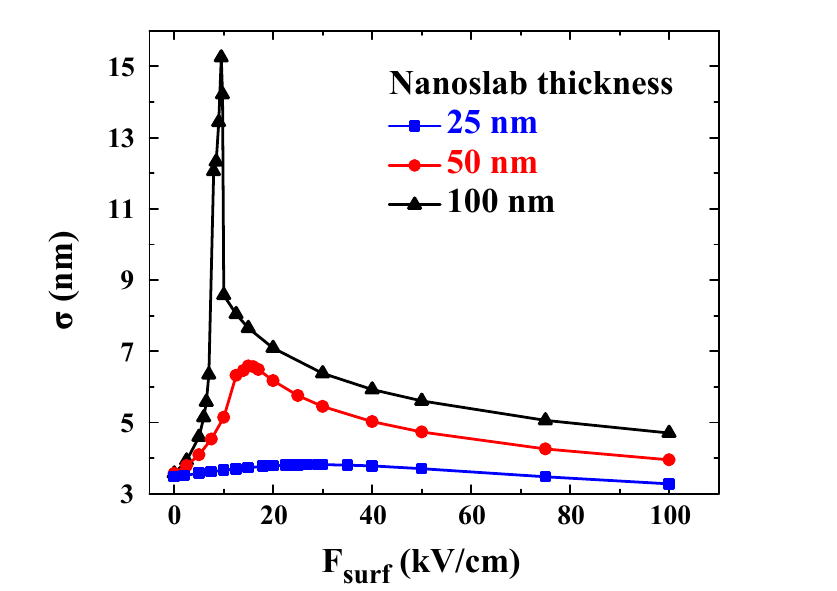}
\caption {(color online): Standard deviation of the electron position $\sigma$ as a function of the surface electric field $F_\text{surf}$ for three nanoslab thicknesses $d$.} 
\label{fig:FigureSigma}
\end{figure} 

\begin{figure}
\includegraphics[scale=1]{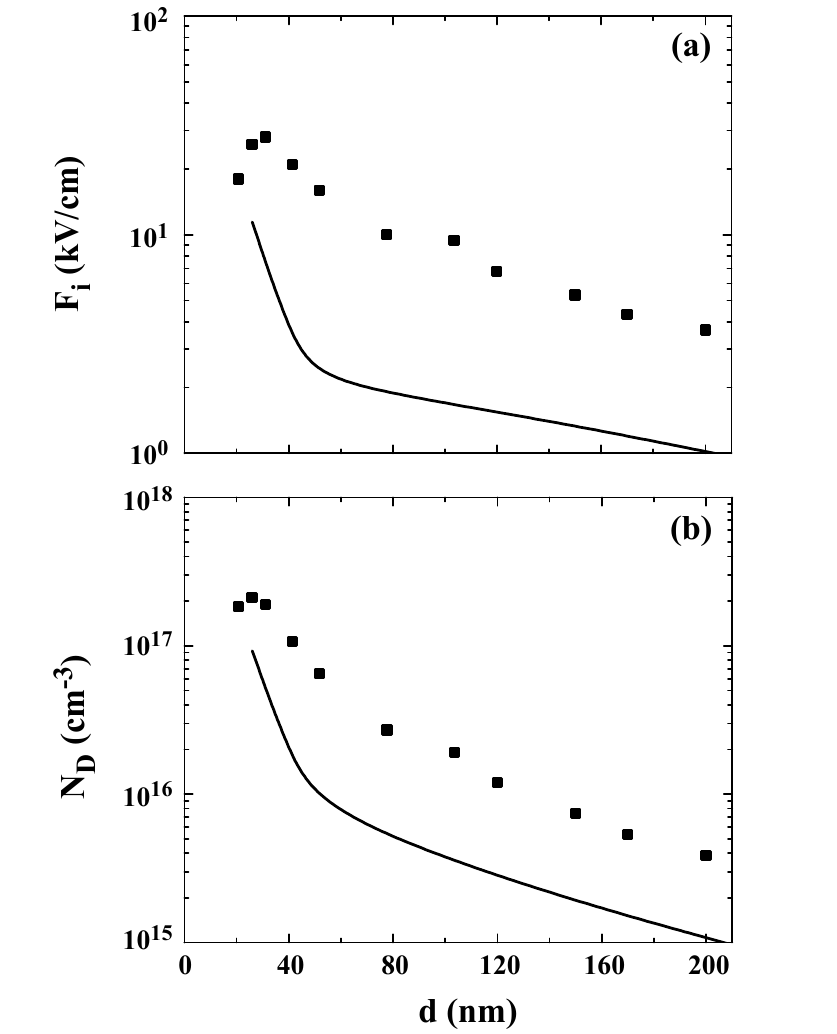}
\caption {: (a) Surface donor ionization field $F_i$ as a function of the nanoslab thickness $d$ (squares). The solid line shows the smallest possible surface electric field for which the surface potential is parabolic for a nanowire with a diameter $\phi=d$. The ionization of surface donors occurs systematically in a regime where the surface potential can be computed with the Poisson equation. (b) Donor concentration N$_\text{D}$ required to induce a surface electric field of magnitude $F_i$ as a function of the slab thickness $d$.} 
\label{fig:FigureFi}
\end{figure} 

Figure~\ref{fig:FigureWavefunction}(a) shows the ground-state wavefunction of an electron bound to a surface donor for a nanoslab  of width $d = 100$~nm. The electron wavefunction is formed from one lobe of a 2$p$-like wavefunction, in agreement with the early works of \citet{Levine1965} and \citet{Satpathy1983}. The binding energy of an electron to such a surface donor is 17.7~meV, significantly larger than the 7~meV predicted by those reports which neglected the dielectric mismatch at the nanoslab surface.\cite{Diarra2007,Pierre2010,Corfdir2012}

Figures~\ref{fig:FigureWavefunction}(b)--\ref{fig:FigureWavefunction}(d) show the evolution of the electron wavefunction for a nanoslab with $d = 100$~nm and for various magnitudes of $F_\text{surf}$. For $F_\text{surf} < 5$~kV/cm, the electron remains bound to the donor atom and the 2$p$ shape of its wavefunction is mostly maintained. For $F_\text{surf} = 7.5$~kV/cm, the electron wavefunction starts to spread towards the center of the nanoslab and its spatial extent increases significantly [cf.\ Fig.~\ref{fig:FigureWavefunction}(c)]. For even larger values of $F_\text{surf}$, the built-in electric fields dominate over the Coulomb interaction between the electron and the donor: the electron wavefunction is centered in the nanoslab, its spreading reduces and its shape corresponds to a Bessel function of the first kind. 

To quantify the change in spatial extent of the electron wavefunction, we utilize the standard deviation of the electron position $\sigma = (\int x^2 |f_e (x)|^2 dx - [\int x |f_e (x)|^2 dx]^2)^{1/2}$ as a function of $F_\text{surf}$ as shown in Fig.~\ref{fig:FigureSigma}. When $\sigma$ is maximum, the parabolic potential that attracts the electron towards the center of the nanoslab counterbalances exactly the attraction exerted by the donor. For $d = 100$~nm and $F_\text{surf} = 0$, $\sigma = 3.7$~nm for a surface donor. The maximum value of $\sigma$ of 15.2~nm is obtained for $F_\text{surf} = 9.5$~kV. For even larger values of $F_\text{surf}$, the electron is delocalized towards the center of the nanoslab. The surface donor is therefore ionized. We note that in the absence of a surface donor, increasing $F_\text{surf}$ leads to a monotonous decrease of $\sigma$.

Figure~\ref{fig:FigureFi}(a) shows the ionization field $F_\text{i}$, defined  as the value of $F_\text{surf}$ that maximizes $\sigma$ for a given $d$, as a function of $d$. The ionization of surface donors occurs for values of $N_D$ for which $V_e$ is parabolic [cf.\ Fig.~\ref{fig:FigureFi}(b)]. Second, the evolution of $F_\text{i}$ with $d$ is not monotonic: as shown in Fig.~\ref{fig:FigureFi}(a), the maximum $F_\text{i}$ is 28~kV/cm when $d = 32$~nm. Note that this ionization field is three times smaller than that expected for donors in the bulk.\cite{Pedros2007} When $d$ is decreased below 32~nm, $F_\text{i}$ decreases due to an enhanced interaction between the electron and the image charges located at the interface of the nanoslab on the opposite side of the surface donor. In contrast, when the $d$ is larger than 32~nm, the interaction of the electron with all-but-the-nearest of its images becomes marginal. Therefore, for a given $F_\text{surf}$, the larger the value of $\phi$, the larger the potential difference between the center and the surface of the nanowire, and the smaller will be the ionization field $F_i$.

\section{Discussion}
\label{sec:Discussion}

As discussed in Section \ref{sec:PL}, the linewidth of the donor-bound exciton transition measured at 10~K from a GaN nanowire ensemble formed at temperatures below 850\,°C is comparatively large, namely, about 1--2 meV [cf.~Fig. \ref{fig:FigurePLandSEM}(a) and Refs.~\onlinecite{Calleja2000,Robins2007,Corfdir2009,Park2009,Brandt2010}]. We have shown in Section \ref{sec:BindingEnergy} that the nanowire surface leads to a distortion of the donor wavefunction and, in particular, to a change in electron binding energy. It is safe to assume that the energy of donor-bound excitons (formed of two electrons and a hole orbiting around a charged donor\cite{Suffczynski1989}) as well as their recombination energy are affected in an analogous way. Therefore, even in the absence of microstrain, the donor-bound exciton transitions is intrinsically broadened due to the random distribution of donors in nanowires.\cite{Corfdir2009,Brandt2010}

The result of our present calculations provides further insight into the optical properties of GaN nanowires. First of all, surface donors in nanowires with $\phi$ of the order of 80--100 nm are ionized when $F_\text{surf}$ is larger than about 10~kV/cm [cf.\ Fig.~\ref{fig:FigureFi}(a)]. This field corresponds to donor concentration of $2 \times 10^{16}$ cm$^{-3}$ [Fig.~\ref{fig:FigureFi}(b)], a rather typical value for GaN in general, and close to  the values reported by \citet{Pfueller2010b} and \citet{Sanford2010} for GaN nanowires in particular. Since the binding energy of donor-bound excitons is a fraction of that of the corresponding donors,\cite{Haynes1960} surface donor-bound excitons are not stable in nanowires for which $F_\text{surf}$ is larger than $F_\text{i}$. Second, the large spatial extent of the wavefunction of an electron or of an exciton bound to a surface donor can also contribute to the instability of these complexes in nanowires. As shown in Fig.~\ref{fig:Summary}, for a nanowire with $\phi = 100$~nm, $F_\text{surf}$ increases from 0.9 to 9~kV/cm when $N_D$ increases from $10^{15}$ to $10^{16}$ cm$^{-3}$. This increase in $F_\text{surf}$ is accompanied by an increase of $\sigma$ from 3.7 to 13.4~nm (cf.\ Figure \ref{fig:FigureSigma}). The volume probed by the electron wavefunction therefore increases by a factor of nearly 50 as compared to bulk-like donors. Consequently,  for GaN nanowires of moderate doping density, the wavefunction of a surface donor-bound electron may certainly probe other impurities, e.\,g., neutral donors located at the core of the nanowire, and bind to them if energetically favorable. Note that this comment also applies to donor-bound excitons, whose wavefunction in bulk material is typically two times larger than that of the electron bound to a neutral donor.\cite{Suffczynski1989} 

\section{Conclusion}
\label{sec:Conclusion}

Our calculations demonstrate that surface donor-bound excitons do not form in nanowires with doping levels exceeding a certain, diameter-dependent threshold. As a consequence, the "intrinsic" broadening of the donor-bound exciton transition in GaN nanowires should abruptly disappear at this threshold for a given nanowire diameter. For a spontaneously formed GaN nanowire ensemble with its comparatively broad diameter distribution, the transition may be a gradual one, but we would certainly expect a reduction of the linewidth of the donor-bound exciton transition upon an increase of the doping level from low (10$^{15}$~cm$^{-3}$) to moderate (10$^{17}$~cm$^{-3}$). An abrupt reduction of the linewidth of the donor-bound exciton transition is precisely what we observe experimentally for our GaN nanowire ensembles as soon as Si incorporation sets in at high growth temperatures. Our calculations thus explain the seemingly conflicting results of an increased donor concentration and a reduced linewidth. Assuming that the minimum linewidth of 400~\textmu eV observed for the (O$^0$,X$_\text{A}$) line of the high-temperature GaN nanowire is free from surface contributions, these results allow us to get a glimpse onto the actual structural perfection of GaN nanowire ensembles. Taking into account the instrumental broadening of 250~\textmu eV, the linewidth observed corresponds to a residual inhomogeneous broadening of about 300~\textmu eV, which is probably due to microstrain generated by the coalescence of adjacent nanowires.\cite{Jenichen2011,Kaganer2012,Brandt2014} In any case, the crystal quality of these spontaneously formed GaN nanowires on Si is clearly on par with that of free-standing GaN.\cite{Paskov2007}

\acknowledgements
The authors would like to thank Timur Flissikowski for a critical reading of the manuscript. They are indebted to Lutz Geelhaar, Holger T. Grahn and Henning Riechert for continuous encouragement and support. Financial support of this work by the Deutsche Forschungsgemeinschaft within SFB 951 is gratefully acknowledged.


%

\end{document}